\documentclass[11pt,a4paper]{article}
\usepackage[hyperref]{acl2020}
\usepackage{times}
\usepackage{latexsym}

\usepackage{graphicx}

\usepackage{microtype}

\aclfinalcopy 
\def\aclpaperid{2199} 

\usepackage{amsmath}
\usepackage{bm}
\usepackage{amsfonts}
\usepackage{multirow}
\usepackage{enumitem}

\DeclareMathSizes{8}{10}{10}{8}
\everymath{\displaystyle}

\title{Neural Temporal Opinion Modelling for Opinion Prediction on Twitter}

 \author{
 Lixing Zhu\textsuperscript{\dag} \qquad Yulan He\textsuperscript{\dag}\thanks{~~Corresponding author} \qquad Deyu Zhou\textsuperscript{\S} \\
 \textsuperscript{\dag}Department of Computer Science, University of Warwick, UK\\
 \textsuperscript{\S}School of Computer Science and Engineering, Key Laboratory of Computer Network\\
 and Information Integration, Ministry of Education, Southeast University, China\\
 {\tt \{Lixing.Zhu,Yulan.He\}@warwick.ac.uk} \qquad {\tt d.zhou@seu.edu.cn}
}

\date{}

\begin{document}

\maketitle

\begin{abstract}
Opinion prediction on Twitter is challenging due to the transient nature of tweet content and neighbourhood context. In this paper, we model users' tweet posting behaviour as a temporal point process to jointly predict the posting time and the stance label of the next tweet given a user's historical tweet sequence and tweets posted by their neighbours. We design a topic-driven attention mechanism to capture the dynamic topic shifts in the neighbourhood context. Experimental results show that the proposed model predicts both the posting time and the stance labels of future tweets more accurately compared to a number of competitive baselines.
\end{abstract}

\section{Introduction}
Social media platforms allow users to express their opinions online towards various subject matters. Despite much progress in sentiment analysis in social media, the prediction of opinions, however, remains challenging. Opinion formation is a complex process. An individual's opinion could be influenced by their own prior belief, their social circles and external factors. Existing studies often assume that socially connected users hold similar opinions. Social network information is integrated with user representations via weighted links and encoded using neural networks with attentions or more recently Graphical Convolutional Networks (GCNs) \cite{chen2016content,li2019encoding}. This strand of work, including \cite{chen2018tracking,zhu2020neural,del2019you}, leverages both the chronological tweet sequence and social networks to predict users' opinions.

The majority of previous work requires a manual segmentation of a tweet sequence into equally-spaced intervals based on either tweet counts or time duration. Models trained on the current interval are used to predict users' opinions in the next interval. However, we argue that such a manual segmentation may not be appropriate since users post tweets at different frequency. Also, the time interval between two consecutively published tweets by a user is important to study the underlying opinion dynamics system and hence should be treated as a random variable.

Inspired by the multivariate Hawkes process \cite{aalen2008survival,du2016recurrent}, we propose to model a user's posting behaviour by a temporal point process that when user $u$ posts a tweet $d$ at time $t$, they need to decide on whether they want to post a new topic/opinion, or post a topic/opinion influenced by past tweets either posted by other users or by themselves. We thus propose a neural temporal opinion model to jointly predict the time when the new post will be published and its associated stance. Instead of using the fixed formulation of the multivariate Hawkes process, the intensity function of the point process is automatically learned by a gated recurrent neural network. In addition, one's neighbourhood context and the topics of their previously published tweets are also taken into account for the prediction of both the posting time and stance of the next tweet.

To the best of our knowledge, this is the first work to exploit the temporal point process for opinion prediction on Twitter. Experimental results on the two Twitter datasets relating to Brexit and US general election show that our proposed model outperforms existing approaches on both stance and posting time prediction.

\begin{figure*}[tp!]
\includegraphics[width=1.0\textwidth]{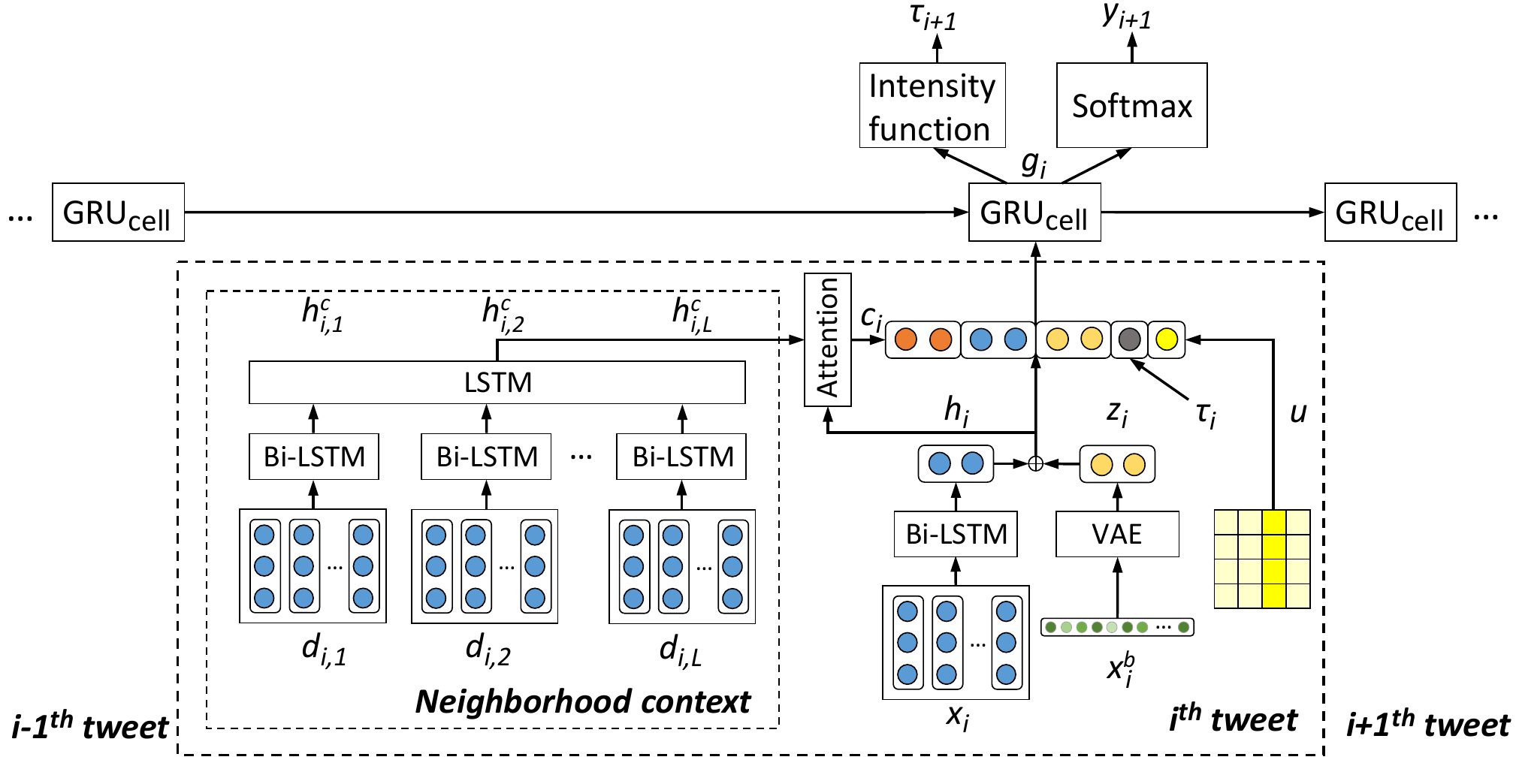}
\caption{Overview of the Neural Temporal Opinion Model.}
\label{fig:1}
\end{figure*}

\section{Methodology}
We present in Figure \ref{fig:1} the overall architecture of our proposed Neural Temporal Opinion Model (NTOM). The input to the model at time step $i$ consists of user's own tweet $x_i$, bag-of-word representation $x^b_i$, time interval $\tau_i$ between the ${i-1}^{th}$ tweet and the $i^{th}$ tweet, user embedding $u$, and neighbours' tweet queue $\lbrace d_{i,1}, d_{i,2}, \dots, d_{i,L} \rbrace$. At first, a Bi-LSTM layer is applied to extract features from input tweets. Then the neighborhood tweets are processed by a stacked Bi-LSTM/LSTM layer for the extraction of neighborhood context, which is fed into an attention module queried by the user's own tweet $h_i$ and topic $z_i$. The output of attention module is concatenated with tweet representation, time interval $\tau_i$, user representation $u$, and topic representation $z_i$, which is encoded from $x^b_i$ via a Variational Autoencoder (VAE). Finally, the combined representation is sent to a GRU cell, whose hidden state participates in computing the intensity function and the softmax function, for the prediction of the posting time interval and the stance label of the next tweet. In the following, we elaborate the model in more details:

\noindent\textbf{Tweet representation: }Words in tweets are mapped to pre-trained word embeddings \cite{baziotis2017datastories}\footnote{\url{https://github.com/cbaziotis/datastories-semeval2017-task4}}, which is specially trained for tweets. Then Bi-LSTM is used to generate the tweet representation.

\noindent\textbf{Topic extraction: }The topic representation $z_i$ in Figure \ref{fig:1} captures the topic focus of
the $i^{th}$ tweet. It is learned by VAE \cite{kingma2014auto}, which approximates the intractable true posterior by optimising the reconstruction error between the generated tweet and the original tweet. Specifically, we convert each tweet to the bag-of-word format weighted by term frequency, $x^b_i$, and feed it to two inference neural networks defined as $f_{\mu_\phi}$ and $f_{\Sigma_\phi}$. These generate mean and variance of a Gaussian distribution from which the latent topic vector $z_i$ is sampled. Then the approximated posterior would be $q_\phi(z_i|x^b_i) = \mathcal{N}(z_i|f_{\mu_\phi}(x^b_i), f_{\Sigma_\phi}(x^b_i))$. To generate the observation $\tilde{x}^b_i$ conditional on the latent topic vector $z_i$, we define the generative network as $p_\varphi(x^b_i|z_i) = \mathcal{N}(x^b_i|f_{\mu_\varphi}(z_i)), f_{\Sigma_\varphi}(z_i))$. The reconstruction loss for the tweet $x^b_i$ is then:
\begin{equation}
\scriptstyle
\mathcal{L}_x = \mathbb{E}_{q_{\phi}(z_i|x^b_i)} \lbrack \mathrm{log}\,p_{\varphi}(x^b_i|z_i)\rbrack \\
-\mathrm{KL}(q_{\phi}(z_i|x^b_i)||p(z_i))
\end{equation}

\noindent\textbf{Neighbourhood Context Attention: }
To capture the influence from the neighbourhood context, we first input the neighbours' recent $L$ tweets to an LSTM in a temporal ascending order. The output of the LSTM is weighed by the attention signals queried by the user's $i^{th}$ tweet and topic:
\begin{gather}
c_i = \sum_{l=1}^{L}\alpha_l h^c_{i,l}\\
\alpha_l \propto \mathrm{exp} (\lbrack h_i^{\mathsf{T}}, z_i^{\mathsf{T}} \rbrack \mathrm{tanh}(W_hh^c_{i,l}+W_zz^c_{i,l}))
\end{gather}

\noindent where $\lbrace h^c_{i,1}, h^c_{i,2}, \dots, h^c_{i,L} \rbrace$ denotes the hidden state output of each tweet $d_{i,l}$ in the neighbourhood context, $z^c_{i,l}$ denotes the associated topic, $h_i$ is the representation of the user's own tweet at time step $i$, and both $W_h$ and $W_z$ are weight matrices.

We use this attention mechanism to align the user's tweet to the most relevant part in the neighbourhood context. Our rationale is that a user would attend to their neighbours' tweets that discuss similar topics. The attention output $c_i$ is then concatenated with a user's own tweet $h_i$ and the extracted topic $z_i$. We further enrich the representation with the elapsed time $\tau_i$ between the posting time of the current tweet and the last posted tweet, and add a randomly initialised user vector $u$ to distinguish the user from others. The final representation is passed to a GRU cell for the joint prediction of the posting time and stance label of the next tweet.

\noindent\textbf{Temporal Point Process: }The goal of NTOM is to forecast the time gap till the next post, together with the stance label. Instead of modelling the time interval value based on regression analysis, we use the GRU \cite{cho2014properties} to simulate the temporal point process.

At each time step, the combined representation $\lbrack c_i, h_i, z_i, \tau_i, u \rbrack$ is input to the GRU cell to iteratively update the hidden state taking into account the influence of previous tweets:
\begin{equation}
g_i = f_{GRU}(g_{i-1}, c_i, h_i, z_i, \tau_i, u)
\end{equation}
where $g_i$ is the hidden state of GRU cell. Given $g_i$, the intensity function is formulated as:
\begin{equation}
    \lambda^* (t) = \lambda(t|\mathcal{H}_i) = \mathrm{exp}(b_\lambda + v_\lambda^\mathsf{T} g_i + w_\lambda t)
\end{equation}
Here, $\mathcal{H}_i$ summarises all the tweet histories up to tweet $i$, $b_\lambda$ denotes the base density level, the term $v_\lambda^\mathsf{T} g_i$ captures the influence from all previous tweets and $w_\lambda t$ denotes the influence from the instant interval. The likelihood that the next tweet will be posted at the next interval $\tau$ given the history is:
\begin{equation}
f^{*}(\tau)=\lambda^* (\tau) \exp \big(-\int_{0}^\tau \lambda^* (t) dt \big)
\end{equation}

The expectation for the occurrence of the next tweet can be estimated using:
\begin{equation}
    \hat{\tau}_{i+1}=\int_{0}^\infty \tau \cdot f^* (\tau)d\tau
\end{equation}

\noindent\textbf{Loss: } We expect the predicted interval to be close to the actual interval as much as possible by minimising the Gaussian penalty function:
\begin{equation}
\mathcal{L}_{time} = \frac{1}{\sigma\sqrt{2\pi}}\exp\big(\frac{-(\tau_{i+1}-\hat{\tau}_{i+1})^2}{2\sigma^2}\big)
\end{equation}

For the stance prediction we employ the cross-entropy loss denoted as $\mathcal{L}_{stan}$. The final objective function is computed as:
\begin{equation}
    \mathcal{L} = \eta \mathcal{L}_x + \beta\mathcal{L}_{time}+\gamma\mathcal{L}_{stan}
\end{equation}
\noindent where $\eta$, $\beta$ and $\gamma$ are coefficients determining the contribution of various loss functions.

\begin{figure}[tp]
\includegraphics[width=0.48\textwidth]{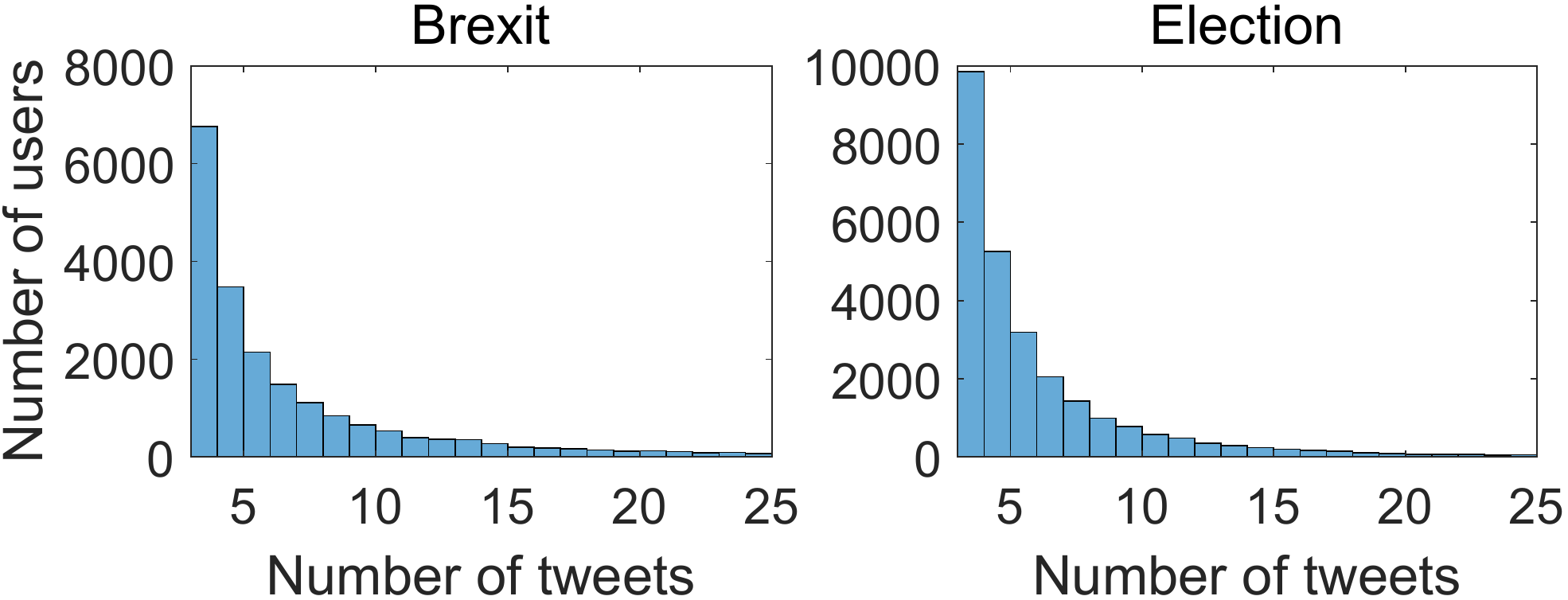}
\caption{Number of users versus number of tweets.}
\label{fig:2}
\end{figure}

\section{Experiments}
\subsection{Setup}
We perform experiments on two publicly available Twitter datasets\footnote{\url{https://github.com/somethingx01/TopicalAttentionBrexit}} \cite{zhu2020neural} on Brexit and US election. The Brexit dataset consists of 363k tweets with 31.6\%/29.3\%/39.1\% supporting/opposing/neutral tweets towards Brexit. The Election dataset consists of 452k tweets with 74.2\%/20.4\%/5.4\% supporting/opposing/neutral tweets towards Trump. We filter out users who posted less than 3 tweets and are left with $20,914$ users in Brexit and $26,965$ users in Election. We plot in Figure \ref{fig:2} the number of users versus the number of tweets and found that over $81.6\%$ users have published fewer than 7 tweets, we therefore set the maximum length of the tweet sequence of each user to 7. For users who have published more than 7 tweets, we split their tweet sequence into multiple training sequences of length 7 with an overlapping window size of 1. For each user, we use 90\% of their tweets for training and 10\% (round up) for testing.

Our settings are $\eta=0.2$, $\beta=0.4$ and $\gamma=0.4$. We set the topic number to 50 and the vocabulary size to 3k for the tweet bag-of-words input to VAE. The mini-batch size is $16$. We use Adam optimizer with learning rate $0.0005$ and learning rate decay $0.9$. The evaluation metrics are accuracy for stance prediction and Mean Squared Error (MSE) for posting time prediction. The results are compared against the following baselines:
\begin{itemize}[leftmargin=*,noitemsep,topsep=0pt,parsep=0pt,partopsep=0pt]
\item[-] CSIM\_W \cite{chen2018tracking} gauges the social influence by an attention mechanism for the prediction of the user sentiment of the next tweet.
\item[-] NOD \cite{zhu2020neural} takes into account the neighborhood context and pre-extracted topics for tweet stance prediction.
\item[-] LING+GAT \cite{del2019you} places a GCN variant over linguistic features to extract node representations. Tweets are aggregated by users for user-level prediction.
\end{itemize}
We also perform ablation study on our model by removing the topic extraction component (NTOM\textsubscript{-VAE}) or removing the neighbourhood context component (NTOM\textsubscript{-context}). In addition, to validate that NTOM does benefit from point process modelling and can better forecast the time and stance of the next tweet, we remove the intensity function (i.e. no Eq. (5)-(7)) and directly use vanilla RNN and its variants including LSTM and GRU to predict the true time interval.
Furthermore, to investigate if is is more beneficial to use GCN to encode the neighbourhood context, we learn tweet representation using GCN\footnote{\url{https://github.com/williamleif/GraphSAGE}} \cite{hamilton2017inductive}, which preserves high-order influence in social networks through convolution. As in \cite{li2019encoding}, we use a 2-hop GCN and denote the variant as NTOM\textsubscript{-GCN}. For the Brexit dataset, MSE is measured in hours, while for the Election dataset it is measured in minutes due to the intensive tweets published within two days.

\begin{table}
\begin{tabular}{|l|c|c|c|c|}
\hline
\multicolumn{1}{|l|}{\multirow{2}{*}{Model}} & \multicolumn{2}{c|}{Brexit} & \multicolumn{2}{c|}{Election} \\
\cline{2-5}
 & \multicolumn{1}{c|}{Acc.} & \multicolumn{1}{c|}{MSE} & \multicolumn{1}{c|}{Acc.} & \multicolumn{1}{c|}{MSE} \\
\hline
CSIM\_W & 0.653& -- &0.656& --\\
NOD & 0.675& -- &0.690& --\\
LING+GAT &0.692& -- &0.704& --\\
\hline
\hline
RNN & 0.636 & 7.81 & 0.659 & 9.62 \\
LSTM & 0.677 & 3.37 & 0.683 & 4.51 \\
GRU &0.691& 2.80 & 0.693 & 3.92 \\
\hline
\hline
NTOM\textsubscript{-VAE} & 0.697&2.67& 0.705 & 4.01\\
NTOM\textsubscript{-context} & 0.665&3.34&0.682& 4.78\\
NTOM\textsubscript{-GCN} & 0.680 & 2.65 & 0.706 & 4.29\\
NTOM & \textbf{0.713} & \textbf{2.59} & \textbf{0.715} & \textbf{3.70}\\
\hline
\end{tabular}
\caption{Stance prediction accuracy and Mean Squared Errors of predicted posting time on the Brexit and Election datasets.}
\label{tab:1}
\end{table}
\subsection{Results}
We report in Table~\ref{tab:1} the stance prediction accuracy and MSE scores of predicted posting time. Compared to baselines, NTOM consistently achieves better performance on both datasets, showing the benefit of modelling the tweet posting sequence as a temporal point process. In the second set of experiments, we study the effect of temporal process modelling. The results verify the benefit of using the intensity function, with at least a 2\% increase in accuracy and 0.2 decrease in MSE compared with vanilla RNN and its variants. In the ablation study, the removal of neighbourhood context component caused the largest performance decline compared to other components, verifying the importance of social influence in opinion prediction. Removing either VAE (for topic extraction) or intensity function (using only GRU) results in slight drops in stance prediction and more noticeable performance gaps in time prediction. It can be also observed that using GCN to model higher-order influence in social networks does not bring any benefits, possibly due to extra noise introduced to the model.

\subsection{Visualisation of Topical Attention}

To investigate the effectiveness of the context attention that is queried by topics, we first select some example topics from the topic-word matrix in VAE. The label of each topic is manually assigned based on its associated top 10 words. Then we display a tweet's topic distribution together with its neighborhood tweets' topic distribution. We also visualize the attention weights assigned to the 3 neighborhood tweets.

\begin{figure}[htp]
\includegraphics[width=0.485\textwidth]{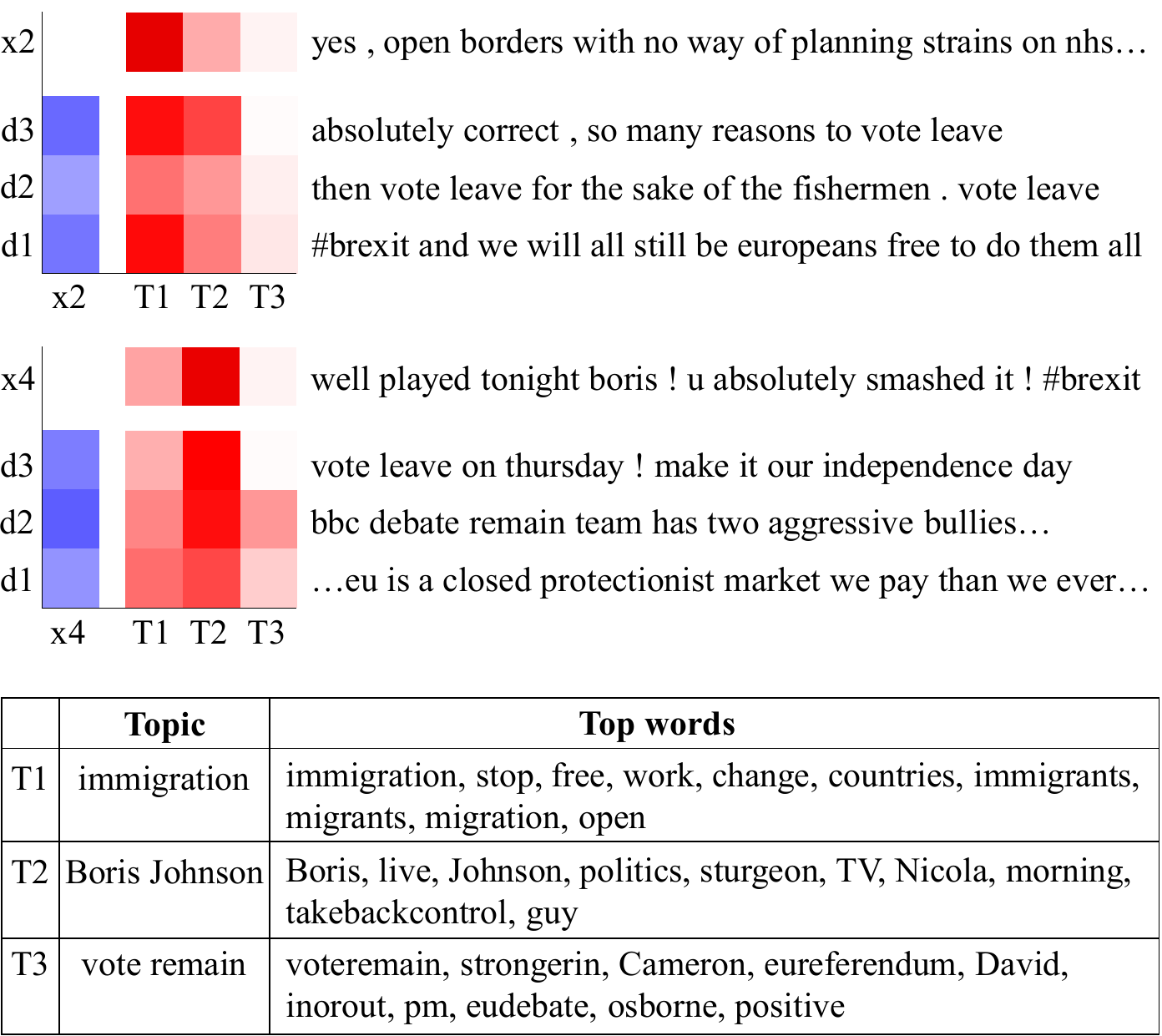}
\caption{Distribution over 3 topics and attention signals on 3 neighborhood tweets, respectively in 2 time steps. Topics are labelled based on the top 10 words.}
\label{fig:3}
\end{figure}

Figure~\ref{fig:3} illustrates the example topics, topic distribution and attention signals towards context tweets. Here, $x_2$ and $x_4$ denote a user's $2^{nd}$ and $4^{th}$ tweets respectively. The most recent 3 neighborhood tweets are denoted as $d_1$, $d_2$, $d_3$. Blue in the leftmost separate column denotes the attention weights, and each row on top of $T1, T2$ and $T3$ denotes the topic distribution. It can be observed that the user's concerned topic shifts  from \textit{immigration} to \textit{Boris Johnson} in 2 time steps. The drift also appears in the neighbour's tweets. Higher attention weights are assigned to the neighbour's tweets which share similar topical distribution as the user. We can thus infer that the topic vector does help select the most relevant  neighborhood tweet.

\section{Related Work}
The prediction of real-time stances on social media is challenging, partly caused by the diversity and fickleness of users~\cite{andrews2019learning}. A line of work mitigated the problem by taking into account the homophily that users are similar to their friends~\cite{mcpherson2001birds,halberstam2016homophily}. For example, \citet{chen2016content} gauged a user's opinion as an aggregated stance of their neighborhood users. \citet{linmei2019heterogeneous} took a step further by exploiting the extracted topics, which discern a user’s focus on neighborhood tweets. Recent advances in this strand also include the application of GCNs, with which the social relationships are leveraged to enrich the user representations~\cite{li2019encoding,del2019you}.

On the other hand, several work has utilized the chronological order of tweets. \citet{chen2018tracking} presented an opinion tracker that predicts a stance every time a user publishes a tweet, whereas~\cite{zhu2020neural} extended the previous work by introducing a topic-dependent attention. \citet{shrestha2019learning} considered diverse social behaviors and jointly forecast them through a hierarchical neural network. However, 
the aforementioned work requires a manual segmentation of a tweet sequence. Furthermore, they are unable to predict when a user will next publish a tweet and what its associated stance is. 
These problems can be addressed using the Hawkes process~\cite{hawkes1971spectra}, 
which has been successfully applied to event tracking~\cite{srijith2017longitudinal}, rumor detection~\cite{lukasik2016hawkes,zubiaga2016stance,alvari2019hawkes} and retweet prediction~\cite{kobayashi2016tideh}. 
A combination of the Hawkes process with recurrent neural networks, called Recurrent Marked Temporal Pointed Process (RMTPP), was proposed to automatically capture the influence of the past events on future events, which shows promising results on geolocation prediction~\cite{du2016recurrent}. Benefiting from the flexibility and scalability of neural networks, several work has been done in this vein including event sequence prediction~\cite{mei2017neural} and failure prediction~\cite{xiao2017modeling}. Our work is partly inspired by RMTPP, but departs from the previous work by jointly 
considering users' social relations and topical attentions for 
stance prediction on social media.

\section{Conclusion}
In this paper, we propose a novel Neural Temporal Opinion Model (NTOM) to address users' changing interest and dynamic social context. We model users' tweet posting behaviour based on a temporal point process 
for the joint prediction of the posting time and stance label of the next tweet. 
Experimental results verify the effectiveness of the model. Furthermore, visualisation of the topics and attention signals shows that NTOM captures the dynamics in the focused topics and contextual attention.

\section*{Acknowledgments}
 This work was funded in part by EPSRC (grant no. EP/T017112/1). LZ is funded by the Chancellor's International Scholarship of the University of Warwick. DZ was partially funded by the National Key Research and Development Program of China (2017YFB1002801) and the National Natural Science Foundation of China (61772132). The authors would also like to thank Akanni Adewoyin for insightful discussions.
\bibliography{acl2020}
\bibliographystyle{acl_natbib}

\end{document}